\documentstyle[psfig,epsf,conf-X,10pt]{article}
\def\listitem{\par \hangindent=50pt\hangafter=1
     $\ $\hbox to 20pt{\hfil $\bullet$ \hfil}}

\def\puncspace{\ifmmode\,\else{\ifcat.\C{\if.\C\else\if,\C\else\if?\C\else%
\if:\C\else\if;\C\else\if-\C\else\if)\C\else\if/\C\else\if]\C\else\if'\C%
\else\space\fi\fi\fi\fi\fi\fi\fi\fi\fi\fi}%
\else\if\empty\C\else\if\space\C\else\space\fi\fi\fi}\fi}
\def\SP{\let\\=\empty\futurelet\C\puncspace}
\def\zmf{$z_{MF}$\SP}

\def\h1{$h^{-1}$\SP}

\def\etal{{\it et al.\/}\ }

\def\lsim{~\rlap{$<$}{\lower 1.0ex\hbox{$\sim$}}}
\def\gsim{~\rlap{$>$}{\lower 1.0ex\hbox{$\sim$}}}
\def\void#1{{}}

\begin{document} 
\small
\heading{%
%
BUILDING A SAMPLE OF DISTANT CLUSTERS OF GALAXIES 
}
\par\medskip\noindent
\author{%
Luiz da Costa $^{1}$, M. Scodeggio$^{1}$, L.F. Olsen$^{2}$,
C. Benoist$^{1}$
}
\address{%
European Southern Observatory, Karl-Schwarzschild Strasse 2,
       D--85748,  Garching
       bei M\"unchen, Germany
}
\address{%
Astronomisk Observatorium, Juliane Maries Vej 30, DK-2100 Copenhagen, 
Denmark 
}
%

\begin{abstract} Candidate clusters of galaxies drawn from the sample
identified from the moderately deep $I-$band data of the ESO Imaging
Survey (EIS), have been used for follow-up optical/infrared imaging
and spectroscopic observations.  The observations were conducted to
assess the nature of these candidates over a large range of
redshifts. Currently, 163 EIS candidates have $(V-I)$ colors, 15 have
$(I-K)$ and 65 cluster fields have been observed
spectroscopically. From a preliminary analysis of these data, we find
that $>$ 65\% of the candidates studied show strong evidence of being
real physical associations, over the redshift range $0.2<z<1.1$. The
evidence in some cases comes directly from spectroscopic measurements,
in others indirectly from the detection of overdensities of objects
with either the same color or the same photometric redshift, or from a
combination of color and spectroscopic information. Preliminary
results also suggest that the redshift derived from the matched-filter
algorithm is a reasonable measure of the cluster's redshift, possibly
overestimating it by $\Delta z\sim0.1$, at least for systems at
$z<0.7$.  Overdensities of red objects have been detected in over 100
candidates, 38 of which with estimated redshifts $>0.6$, and six
candidates in the interval $0.45<z<0.81$ have either been identified
directly from measured redshifts or have been confirmed by the
measurement of at least one redshift for galaxies located along a
red-sequence typical of cluster early-type galaxies. Lastly, five
candidates among those already observed in the infrared have $(I-Ks)$
colors consistent with them being in the redshift interval
$0.8<z<1.1$.  The sample of "confirmed" clusters, already the largest
of its kind in the southern hemisphere, will be further enlarged by
ongoing observations.

\end{abstract} 

\section{Introduction}

The recent discovery of apparently massive and relaxed clusters of
galaxies at redshifts $z\gsim0.5$ offers a unique opportunity to study
the evolution of gravitationally bound systems over an extended
look-back time. Moreover, if such systems are proven to be massive,
especially those at $z\gsim0.8$, their sheer existence can impose
stringent constraints on viable cosmological
models~\cite{bahcall}~\cite{eke}.  In addition, a large sample of
confirmed clusters, spanning a broad redshift range, is of great
interest for constraining models of formation and evolution of
early-type galaxies and large-scale structure, and for the selection
of targets in different redshift intervals for a variety of other
applications. While the existence of high-redshift clusters is by now
generally accepted, current samples of spectroscopically confirmed
clusters, predominantly $X$-ray selected, are small and unlikely to
grow in the near future, at least not until deep observations with
Chandra and XMM are carried out and analyzed.  This is particularly
true at the highest redshifts ($z\sim1$), where only a handful of
systems have been confirmed so
far~\cite{rosatia}~\cite{stanford}~\cite{rosatib}.

An alternative approach is to use moderately deep $I$-band surveys to
identify overdensities in the projected distribution of
galaxies. Surveys suitable for this purpose include the Palomar
Distant Cluster Survey~\cite{postman}, covering over 5 square degrees
in the northern hemisphere, and the ESO Imaging Survey
(EIS)~\cite{olsena}~\cite{olsenb}~\cite{scodeggioa}, covering about 17
square degrees in the southern hemisphere. The main advantage of
optical searches is that large areas of the sky can be covered using
wide-field imagers. However, the yield of physical associations from
such samples is largely undetermined, as projection, among other
effects, may lead to spurious detections. Therefore, assessing the
success rate of the identification of real systems at different
redshifts is an essential step in determining the possible role, and
the best design, of ground-based wide-angle surveys in building a
large sample of confirmed clusters over a large redshift range
suitable for evolutionary studies.

In this contribution, we describe the ongoing effort to establish the
yield of real clusters from the EIS distant cluster candidate list. To
this end we combine the results obtained from follow-up observations
by different groups. This ongoing work includes optical/infrared
imaging and spectroscopic observations, which are used to provide
either direct or circumstantial evidence supporting the identification
of real systems. The confirmation of a cluster candidate requires one
or more of the following conditions: the identification of concordant
redshifts from spectroscopic observations; the existence of a
distinctive red-sequence in optical/infrared color-magnitude diagrams;
the detection of a statistically significant concentration of galaxies
with similar colors or photometric redshifts.

\section{Analysis of Photometric Data}

The original EIS catalog of cluster candidates was constructed by
applying a matched-filter algorithm~\cite{postman} to galaxy catalogs
extracted from the EIS $I-$band survey
images~\cite{noninoa}~\cite{prandoni}~\cite{benoist}. The survey
covers four patches of the sky spread over the right ascension range
$22^h\lsim \alpha \lsim~10^h$ and the object catalogs reach a limiting
magnitude of $I_{AB}\sim23$. Details about these data, the location of
the patches and the EIS cluster candidate catalog can be found in the
original papers~\cite{olsena}~\cite{olsenb}~\cite{scodeggioa}. More
recently, most of the area covered by the $I$-band survey has been
observed in $B$ and $V$ as part of the public survey being conducted
by EIS using the wide-field camera now available at the MPG/ESO 2.2m
telescope at La Silla~\cite{dacosta}~\cite{noninob}. A summary of the
data currently available, combining these new data with those in the
$B$ and $V$ passbands of the original survey~\cite{prandoni}, is shown
in Table~\ref{tab:coverage}. The table gives, for each EIS patch, the
total area covered in each passband. The survey is still ongoing and
it is expected to cover roughly 14 square degrees in $BVI$. The
available data provide color information for a significant fraction of
the EIS cluster candidates. Nearly all candidates already have $BI$
and over half of them have $BVI$ data.

\begin{table}
\begin{center}
\caption {\bf Wide-Angle Survey (square degrees)}
 \vskip 0.5cm
\begin{tabular}{lrrrrr} \hline\hline
\label{tab:coverage}
Patch   & U  &  $B$  &   $V$   & $I$ \\
\hline
A    & - & 1.0 & 1.2  &   3.2    \\
B   & 1.25 & 1.5 & 1.5  &  1.6 \\
C   &- &    6.0 &  -   &     6.0  \\
D   &- &    6.0   & 6.0   &  6.0   \\
\hline
Total & 1.25 & 14.5 & 8.7  & 16.8 \\
\hline\hline
\end{tabular}
\end{center}
\end{table}

Also available are data obtained by different groups from
optical/infrared observations of EIS cluster candidates with
$z>0.5$. These data were obtained at the NTT~\cite{scodeggiob}
(infrared observations for 15 clusters) and the Danish 1.5m telescope
on La Silla, and the Nordic Optical Telescope~\cite{olsen} ($V$ and
$R$ observations for 50 clusters). These observations are still
ongoing, which will enable doubling the sample within the next year.

\subsection{The red-sequence of early-type galaxies}

Even though there is no substitute for direct spectroscopic
confirmation of candidate cluster of galaxies, these observations are
extremely time consuming. In fact, for this type of work it is not
only important to have a high-confidence that the selected candidates
are real associations but also to be able to select individual
galaxies likely to be cluster members.

One way of pre-selecting candidates is to use the available
multi-color data to search for galaxies with red colors corresponding
to the cluster early-type galaxy population. Such red-sequences have
been detected over a broad redshift range using optical and
optical/infrared colors for low and high redshift clusters,
respectively~\cite{kodama}~\cite{stanford}. As shown below, from the
data we have, optical colors are useful for $z\lsim0.7$, while
optical/infrared data are required to extend the redshift range beyond
this limit.

In order to find supporting evidence that the EIS candidates are
physical associations, those with available data in the $V$ and $I$
filters have been analyzed to search for the presence of a
red-sequence. This has been done either by direct inspection of the
$(V-I)-I$ color-magnitude diagrams where, at least for clusters with
$z\lsim0.4$, a red-sequence appears as a distinctive
feature\cite{olsenb}, or by evaluating the statistical
significance of spatial concentrations of galaxies in different color
slices. The significance of a detection is determined using mock
catalogs containing the same number of objects as the data, as
described in more detail in forthcoming
papers~\cite{olsen}~\cite{scodeggiob}.

\begin{figure}
\mbox{\epsfxsize=10cm\epsffile{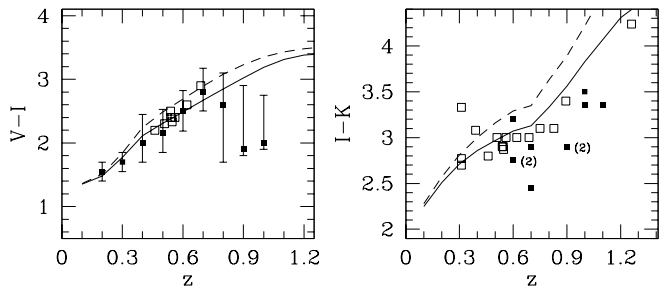}}
\caption{The relation between the median (left panel) or individual
(right panel) color and redshift for the clusters detected by the the
color-slice analysis (solid squares). Also shown are model predictions
for a non-evolving elliptical galaxy (dashed line) and a passively
evolving one (solid line), and colors of cluster galaxies in
spectroscopically confirmed clusters (open squares). For the EIS
clusters the redshift is assumed to be that estimated by the
matched-filter algorithm.}
\label{fig:vizmf}
\end{figure}

Considering the 105 detections at the 95\% confidence level obtained
from the color-slice analysis we show, in the left panel of
Figure~\ref{fig:vizmf}, the relation between the $(V-I)$ color
characterizing the detected concentrations and the redshift. For the
EIS candidates we take the redshift to be $z_{MF}$ as given by the
matched-filter. For $z\lsim0.7$ we show the median and the quartiles
of the color distribution of the detected red galaxy
concentrations. At higher redshifts, the error bars represent the full
range of colors covered by the detections using deeper $V$ data. Also
shown are curves corresponding to the expected variation of the
$(V-I$) color with redshift for a non-evolving elliptical galaxy and a
passively evolving one, assuming a formation epoch at $z>4$, a single
burst of star formation, and a subsequent evolution in a low-density
cosmological model. The colors of early-type galaxies along the
red-sequence of spectroscopically confirmed
clusters are also shown~\cite{stanford}~\cite{kodama}.

From the figure we find that \zmf is by and large a fair indicator of
the cluster redshift, albeit the large scatter, predicting colors
consistent with the models and empirical data.  The scatter in color
is to a large extent due to the uncertainties in the detection
procedure, errors in the galaxy colors and the uncertainties
associated with \zmf. The consistency between the medians of the EIS
candidate colors, the models and the empirical data up to $z\lsim0.7$
is remarkable.  For larger redshifts most cluster candidates are bluer
than expected for their estimated redshift. From the visual inspection
of these candidates (in different passbands) we find that the blue
colors are either due to contamination by a foreground cluster or they
simply reflect the fact that galaxies in high-redshift clusters
are not detected in the $V$-band.  These results show that the effectiveness
of the $(V-I)$ color in confirming cluster candidates is in practice
limited to systems with $z\lsim0.7$, with deeper $V$ data providing
little additional leverage.

We have also examined the CM-diagrams based on the available
optical/infrared data for 15 clusters with $z_{MF}\gsim0.6$. From the
analysis of $(I-Ks)$ color slices, significant overdensities are
detected near the nominal center of the candidates in 10 of the
observed fields. The same systems are also identified in $(J-Ks)$, but
the weak dependence of this color on the redshift yields no
independent constraint on the redshift. For the $>$95\% confidence
level detections, the colors range from about 2.4 to 3.6 for $(I-Ks)$
and 1.6 to 2.1 in $(J-Ks)$. The dependence of the $(I-Ks)$ color of
cluster galaxies on redshift is shown in the right panel of
Figure~\ref{fig:vizmf}. For comparison we also show the expected
dependence of the $(I-Ks)$ color of ellipticals on redshift, for the
same models presented earlier, and the color-redshift relation for
clusters observed
spectroscopically~\cite{stanford}~\cite{rosatib}. Given the small
number of candidates considered, we show the color of each detected
candidate. The number in parenthesis indicates cases of overlap.
While for low redshifts this color is a poor redshift indicator, it
does give some leverage for redshifts $z\gsim0.7$. Note, in
particular, that the colors of the EIS candidates are at least
consistent with their estimated redshifts and the general trend of the
theoretical models, with a few cases having colors consistent with
$z\sim1$.

\subsection{Photometric Redshifts}

\begin{figure}
\mbox{\epsfxsize=10cm \epsffile{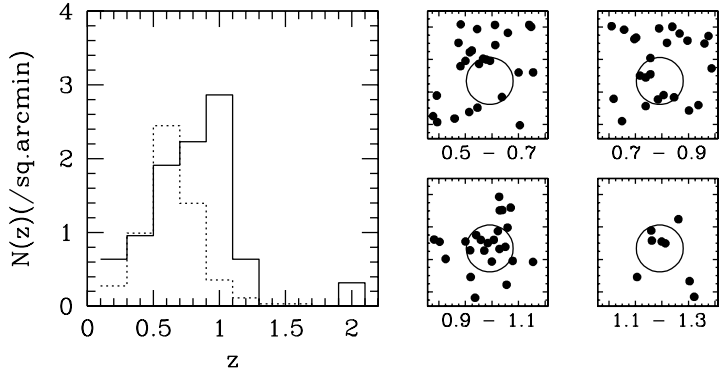}}
\caption{Left panel shows the photometric-redshift distribution for
galaxies with available multi-color data (solid histogram) in one of
the cluster fields. Also shown is the redshift distribution for
background galaxies (dotted histogram). The four panels in the right
show the projected distribution of galaxies in 0.2 redshift bins. The
field is roughly $5\times5$~arcmin and the circle represents the
expected size of a cluster at this redshift.}
\label{fig:map} \end{figure}

An alternative approach for testing the reality of candidate clusters
is to resort to photometric redshifts. The main advantages are that
the method makes optimal use of the multi-color data and that it
enables one to consider not only early-type galaxies but the whole
range of SEDs.  Tests carried out with EIS-deep data have
demonstrated~\cite{benoistb} that the minimal set of filters required
to obtain reasonable estimates of redshifts, taking into account the
redshift range of interest and the limiting magnitudes in the various
colors, is composed of $BVIJK$. Currently, we have such a five-band
coverage for 14~clusters. Photometric redshifts have been
assigned\cite{arnouts}\cite{benoistb}, for all galaxies brighter than
$I=22$ within a $5\times 5$ arcmin field of the SOFI detector.  The
redshift distribution in the field of one of the candidates considered
is shown in Figure~\ref{fig:map}, where it is compared to the
background redshift distribution as measured from all the available
fields. As can be seen the background distribution peaks at
$z\sim0.6$, while for the field considered the redshift distribution
peaks at $z\sim1$.  To evaluate the significance of these peaks we
have resorted to the same method employed above in analyzing color
slices, but considering redshift slices.  We find that in six out of
the seven cases analyzed so far a concentration is
detected. Figure~\ref{fig:map} illustrates the distribution of
galaxies brighter than $I=22$ in photometric redshift slices 0.2 in
width, for one of the cases considered. An overdensity of galaxies is
clearly seen in the redshift interval 0.9-1.1, which coincides with
the position obtained by employing the matched-filter algorithm which,
by the way, also predicts $z_{MF}\sim1$. More importantly, from these
redshift estimates individual galaxies can be selected for
spectroscopic observations over a much larger area than that covered
by early-type galaxies detected using a single color criterion.

\section{Analysis of Spectroscopic Data}

The spectroscopic data are from observations conducted at the 3.6m
telescope at La Silla using the EFOSC2
spectrograph~\cite{ramella}~\cite{dacostaa} and from a single 3 hours
test exposure using the 2dF spectrograph at AAT~\cite{colless}. The
EFOSC2 observations yielded 69 redshifts in the field of seven EIS
cluster candidates selected in the redshift interval $0.5\leq
z_{MF}\leq0.7$. The observations conducted at the 2dF cover the
vicinity of 58 cluster candidates in Patch~D, nearly all of the
clusters over about two-thirds of the total area of the $I$-band
survey, irrespective of their estimated redshift. A total of 340
galaxies were observed yielding 197 redshifts.

A likelihood analysis of the redshift distribution of the galaxies
observed with EFOSC2 was performed, based on a comparison with results
obtained from random re-sampling of the CFRS redshift
distribution. This analysis has shown that four of the seven systems
correspond to a real system in redshift space with a probability
$>95$\%~\cite{ramella}. Two of them have a mean redshift in agreement
with $z_{MF}$, while the other two have  significantly smaller
redshifts.  However, it is important to note that the redshift
distribution in each cluster field is under-sampled and the
identification of bound systems from the redshift distribution alone
may not suffice for a conclusive determination about the nature of the
system and its redshift. In this case it is important to take into
consideration other constraints such as the detection of overdensities
in color slices, the spatial distribution of galaxies with measured
redshifts and their brightness.  Therefore, we supplement the above
spectroscopically confirmed sample with systems for which an
overdensity of red galaxies is detected and at least one galaxy along
this red-sequence has a measured redshift.  When these cases are
considered three additional clusters are identified with redshifts
much closer to those estimated by the matched-filter.  In one case,
even though only one redshift is available, it corresponds to the
brightest galaxy in the field. Moreover, this galaxy lies on a clearly
visible red-sequence, consistent with the measured redshift
($z=0.81$)~\cite{dacostaa}. Furthermore, in one case the low redshift
assigned by the likelihood analysis is very likely due to the
superposition of a nearby system, since one of the measured redshifts
is consistent with that inferred from the color analysis and with that
estimated by the matched-filter. When combined, all the information
available suggests that all seven clusters can be confirmed and, with
exception of one, all observed candidates have $z\gsim0.5$.  Clearly,
denser sampling of these fields is required to resolve some of these
ambiguities.

While no detailed analysis has been carried out of the accumulated 2dF
data, the completeness of this preliminary survey as a function of
magnitude immediately indicates that only clusters with $z\lsim0.4$
can be identified from the redshifts. From a preliminary inspection of
the redshift distribution, identifying cases with two or more
concordant redshifts and, as discussed above, by examining measured
redshifts and color detections, we find that for 22 candidates,
corresponding to 60\% of those with $z<0.4$, it is possible to
tentatively assign a spectroscopic redshift to the cluster.

\begin{figure}
\mbox{\epsfxsize=10cm \epsffile{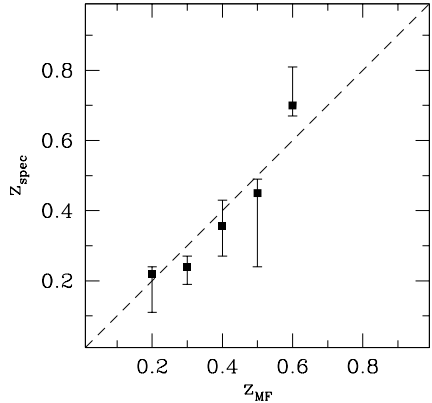}}
\caption{The relation between measured and estimated redshift for the
clusters with spectroscopically assigned redshifts.}
\label{fig:zspeczmf}
\end{figure}

Combining all the available spectroscopic data we can assess how well
the redshift estimated by the matched-filter reflects the true
redshift.  In Figure~\ref{fig:zspeczmf}, we compare the median of the
measured redshift distribution to the original \zmf. Error bars again
represent the full range of the distributions. The results show that
the matched-filter redshift is indeed a fair indicator of the cluster
redshift at least out \zmf $\sim0.7$. On average it overestimates the
redshift by $\Delta z \sim 0.1$, comparable to the observed scatter.
Even though these preliminary results are admittedly based on a small
sample they are encouraging since most of the candidate clusters
observed were drawn directly from the matched-filter list without any
prior information.

\section{Conclusions}

In this paper we have investigated the reality of the $I$-band
selected EIS cluster candidates over an extended redshift
baseline. Our main findings are:

\begin{enumerate}

\item For most cluster candidates with $z_{MF}\lsim 0.4$, red-sequences
can be clearly seen in the $(V-I)-I$ CM diagram leaving little doubt
that they are real clusters.

\item The success rate of detecting concentrations of red galaxies is
$\gsim$ 70\% out to \zmf $\sim$ 0.8, dropping to about 40\% at \zmf
$\sim$ 1.

\item There is a clear correlation between colors and the redshift
estimated by the matched-filter algorithm.  This correlation is seen up
to \zmf$\lsim 0.7$ when the $(V-I)$ color is used and up to \zmf$\sim1$
when $(I-Ks)$ is considered.  The color-redshift relation is consistent
with models of stellar population evolution and/or the color measured for
elliptical galaxies in clusters with measured redshifts.

\item Preliminary spectroscopic observations show that the measured
redshifts are consistent with those estimated by the 
matched-filter algorithm

\end{enumerate}

With the estimated success rate from spectroscopy at $z<0.6$ and from
imaging for larger redshifts we have demonstrated that large samples
of clusters are within reach comprising low ($z\gsim0.2$) to very
high-redshift clusters ($z\sim1$). Such a sample would be ideal for
evolutionary studies of galaxies and large-scale structures.  The EIS
cluster catalog being compiled will also be of great value for many
other applications and in support to VLT observations as well as
targets for follow-up observations with Chandra and XMM.  There is
currently no comparable sample in the southern hemisphere.

\acknowledgements{We would like to thank all of our collaborators for
allowing us to report on the data prior to their publication, in
particular S. Arnouts, S. Bardelli, A. Biviano, M. Colless,
H. J{\o}rgensen, M. Ramella and R. Saglia.  Special thanks to all
members of the EIS team for their effort in keeping the data flowing,
in particular to M. Nonino nd R. Rengelink for their efforts in
processing the WFI and SOFI data, respectively. We would also like to
thank A. Renzini for many useful discussions. Finally, LndC would like
to thank the Local organizers, in particular Manolis Plionis, for a
great meeting in a memorable location.}

\vfill
\end{document}